\documentclass[usenatbib,twocolumn]{mnras}

\usepackage[english]{babel}
\usepackage[utf8x]{inputenc}
\usepackage[T1]{fontenc}

\usepackage{subfigure}
\usepackage{amsmath}
\usepackage{graphicx}
\usepackage[colorinlistoftodos]{todonotes}
\usepackage{amssymb}
\usepackage{soul}

\usepackage[T1]{fontenc}
\usepackage{amsbsy}
\newcommand{\be}{\begin{equation}}
\newcommand{\ee}{\end{equation}}

\def\bi#1{\hbox{\boldmath{$#1$}}}

\newcommand{\tTheta}{\tilde{\Theta}}
\newcommand{\bl}{{\ensuremath{\boldsymbol\ell}}}
\newcommand\nhat{\hat{\mathbf n}}
\def\L{{\bf L}}
\def\l{{\bf  \ell}}

\def\d{{\bf d}}

\title{Reconstructing Small Scale Lenses from the Cosmic Microwave Background Temperature Fluctuations}

\author[B. Horowitz et al.]{
Benjamin Horowitz,$^{1,2}$
Simone Ferraro,$^{1,3,4}$
Blake D. Sherwin,$^{5,1}$
\\
$^{1}$Berkeley Center for Cosmological Physics, University of California,
Berkeley, CA 94720\\
$^{2}$Department of Physics,
University of California, Berkeley, CA, USA 94720\\
$^{3}$Department of Astronomy,
University of California, Berkeley, CA, USA 94720\\
$^{4}$Miller Institute for Basic Research in Science, University of California, Berkeley, CA, 94720 USA\\
$^{5}$Department of Applied Mathematics and Theoretical Physics, University of Cambridge, Wilberforce Road, Cambridge CB3 0WA}

\begin{document}
\maketitle

\begin{abstract}
Cosmic Microwave Background (CMB) lensing is a powerful probe of the matter distribution in the Universe. The standard quadratic estimator, which is typically used to measure the lensing signal, is known to be suboptimal for low-noise polarization data from next-generation experiments. In this paper we explain why the quadratic estimator will also be suboptimal for measuring lensing on very small scales, even for measurements in temperature where this estimator typically performs well. Though maximum likelihood methods could be implemented to improve performance, we explore a much simpler solution, revisiting a previously proposed method to measure lensing which involves a direct inversion of the background gradient. An important application of this simple formalism is the measurement of cluster masses with CMB lensing. We find that directly applying a gradient inversion matched filter to simulated lensed images of the CMB can tighten constraints on cluster masses compared to the quadratic estimator. While the difference is not relevant for existing surveys, for future surveys it can translate to significant improvements in mass calibration for distant clusters, where galaxy lensing calibration is ineffective due to the lack of enough resolved background galaxies. Improvements can be as large as $\sim 50\%$ for a cluster at $z = 2$ and a next-generation CMB experiment with 1$\mu$K-arcmin noise, and over an order of magnitude for lower noise levels.  For future surveys, this simple matched-filter or gradient inversion method approaches the performance of maximum likelihood methods, at a fraction of the computational cost.
\end{abstract}


\section{Introduction}

Weak gravitational lensing of the Cosmic Microwave Background (CMB) by the large-scale structure of the Universe produces a signature in the CMB anisotropies in both temperature and polarization and provides a unique probe of the matter distribution out to high redshift (\cite{seljak1995gravitational}). Lensing induces statistical anisotropies in the observed CMB, through which the mass distribution can be probed. The most common method of lensing reconstruction relies on the quadratic estimator (QE) of Hu and Okamoto (\cite{hu2002mass}).
Previous work (\cite{hirata2003analyzing,carron2017maximum}) has shown that, since the QE is only an approximation to the maximum likelihood solution, significant improvements are possible in polarization at low noise levels, usually at the cost of a much greater computational complexity.

In this paper, we revisit lensing reconstruction on very small scales, in a regime where the primary CMB fluctuations have been Silk damped and are negligible (i.e. $\ell \gtrsim 4000$). We will first show that the quadratic estimator can be very suboptimal in this regime as well, since it is limited by the cosmic variance of the long-wavelength background gradient mode, while a true maximum likelihood solution should be able to reconstruct small lenses arbitrarily well, given small enough instrumental noise and residual foreground levels.  We then show that the statistical power of the QE is further reduced by the fact that the errors become highly correlated, so that the utility of measuring more modes is reduced when the proper covariance between them is taken into account. Finally, we show that a simple gradient inversion matched filtering approach, as proposed by \cite{seljak2000lensing}, is close to optimal for reconstruction of very small scale lenses, and that it is not limited by cosmic variance, thus in principle allowing arbitrarily large improvements over the QE (for small enough noise).

One application of this formalism with cosmological significance is cluster mass determination though CMB lensing, or ``cluster lensing''.
The number counts of galaxy clusters are a key prediction of the $\Lambda$-CDM model which is sensitive to the overall matter density, $\Omega_m$, as well as the amplitude of density fluctuations, $\sigma_8$, and hence the neutrino mass $\Sigma M_\nu$. A key limiting factor of using galaxy clusters for cosmological studies is their mass calibration, which is difficult to obtain (\cite{rozo2013cluster}). Since it isn't directly observable, cosmologists are left to study different proxies of mass, such as X-ray luminosity or Sunyaev-Zel'dovich signal, and correlate them with gravitational effects of the cluster, such as galaxy-cluster lensing. The uncertainty in the calibration of these relations now limits the constraining power of cluster counts as a cosmological probe (\cite{mantz2014cosmology}).

While low-redshift massive clusters can be accurately weighed by using lensing of background galaxies along the line of sight (\cite{von2014weighingI,Madhavacheril:2017onh}), this approach fails for more distant clusters (at $z \gtrsim 1$) because of the lack of enough resolved galaxies on the background. In addition, uncertainties in the photometric redshift distribution of source galaxies, which is challenging to determine precisely at high redshift, can lead to errors in the mass determination.

Meanwhile, the CMB provides a source at known redshift ($z \approx 1100$) and with well studied statistical properties. By looking at the lensing of the CMB by these galaxy clusters, we can avoid some of the shortcomings of galaxy lensing surveys and probe more distant clusters  (\cite{holder2004gravitational,yoo2008improved,yoo2010lensing, melin2015measuring}). While development of the theory behind these measurements dates back almost two decades (\cite{zaldarriaga1998gravitational, seljak2000lensing,lewis2006cluster}), it is only fairly recently that this lensing has been detected at high statistical significance (\cite{melin2015measuring,madhavacheril2015evidence,baxter2015measurement})

These detections have used a modified version of the QE (\cite{hu2007cluster}) which relies on reconstructing the lensing convergence $\kappa$, and then applying a matched filter (\cite{melin2015measuring}) on the resulting convergence map. However, as discussed in (\cite{yoo2008improved}), QE obtains unbiased and optimal results only in the limit of
no gravitational lensing and becomes progressively more biased and sub-optimal as the gradient in lensing potential increases. Clusters found in SZ samples have particularly strong potential gradients due to their large mass and pose a particularly serious problem for quadratic estimator techniques. The work by (\cite{baxter2015measurement}) employs a maximum-likelihood approach and should be immune to the discussion above, but is more computationally expensive.

Revisiting the work of \cite{seljak2000lensing} in the context of upcoming CMB experiments such as CMB S4 (\cite{cmbs4}), we discuss how an unbiased estimate for the mass of a small spherically symmetric lens (such as an idealized cluster) can be obtained with a simple matched filter of the temperature and polarization maps. In the low-noise regime, we show that these estimates are close to the maximum likelihood solution, and considerably better than the QE. Reconstructing the lensing potential on small scales is particularly important as upcoming CMB experiments can potentially reach sensitivities of $\sim$ 1 $\mu$K-arcmin and 1 arcmin beams size. With this increase in instrumental sensitivity, there should be corresponding increases in cluster mass calibration and cosmological constraints (\cite{louis2016calibrating, Madhavacheril:2017onh}). 

In Section \ref{sec_Form}, we briefly review weak lensing of the CMB, explore the small-scale limit of the quadratic estimator and explain the matched filter approach, as well as our assumptions about the cluster mass profile. In Section \ref{sec_mf}, we introduce a matched filter technique we use directly on the temperature map. In Section \ref{sec_idealized}, we implement our technique on realistic lensed CMB maps and compare against the quadratic estimator, finding improved performance at low noise levels and high redshifts. In Section \ref{sec_comparison}, we compare the results of the matched filter estimate to other lensing reconstruction techniques. Finally, in Section \ref{sec_sum}, we will summarize our results and look at prospects for applying it to future experiments. 

For our analysis we will assume the flat $\Lambda$CDM Planck 2013 cosmology (\cite{collaboration2013planck}), with $H_0=67.8$  km (Mpc s)$^{-1}$, and $\Omega_m=0.307$. We will assume the Born approximation, and ignore field rotation/multiple-lens effects throughout this work.

\section{Formalism}
\label{sec_Form}

\subsection{Effect of Cluster Lensing on the CMB}
\label{sec_effects}

\begin{figure}
  \centering  \includegraphics[trim=3cm 3cm 4cm 4cm,width=0.40\textwidth]{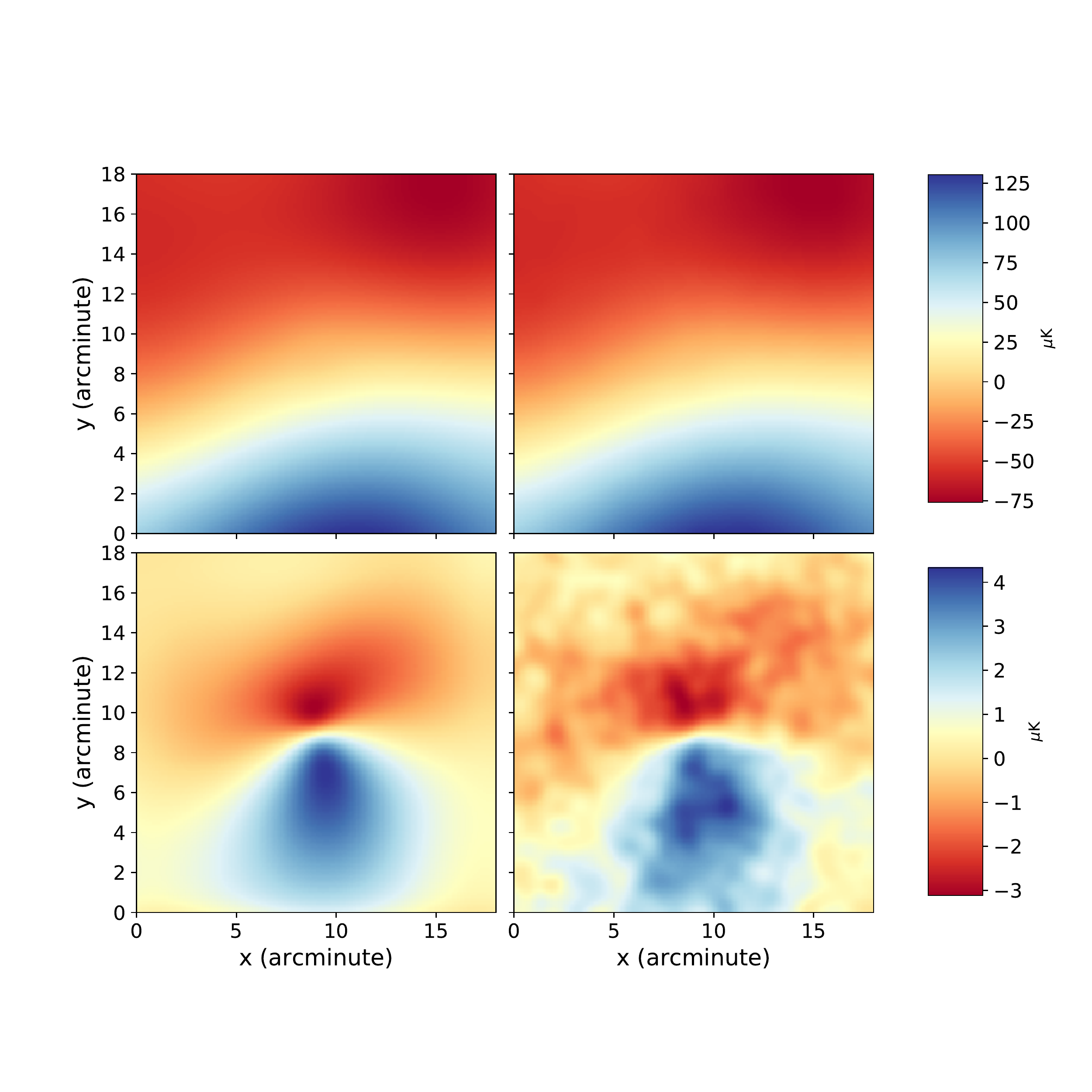}
    \caption{Top: Simulated images of CMB lensing caused by a $2\times 10^{14}$ $M_\odot$ NFW cluster at $z=0.7$. Bottom: Unlensed image subtracted out. Left: Beam alone, Right: Beam and 1 $\mu$K-arcmin noise.} 
    \label{fig_sims2x2}
\end{figure}

Here we briefly review the effects of gravitational lensing on the CMB; for a more complete review see (\cite{lewis2006weak}). The observed CMB anisotropies can be described by their temperature fluctuations as a function of direction $\nhat$, $\Theta(\nhat)$, as well as two Stokes parameters describing their polarization $Q(\nhat)$ and $U(\nhat)$. Since the effect of lensing is a simple coordinate remapping, the observed (i.e. lensed) field, $(\tilde{\Theta},\tilde{Q},\tilde{U})$, can be related to the primary fields, $(\Theta,Q,U)$, through a deflection angle field $\d(\nhat) = \nabla \phi$, where $\phi$ is the lensing potential, by 

\begin{subequations}
\begin{align}
        \tTheta(\nhat)&=\Theta(\nhat+\nabla \phi)) \approx {\Theta}(\nhat) + \nabla \phi \cdot \nabla {\Theta}(\nhat) + \ldots \\
        \tilde{Q}(\nhat)&=Q(\nhat+\nabla \phi)) \approx Q(\nhat) + \nabla \phi \cdot \nabla Q (\nhat) + \ldots\\
        \tilde{U}(\nhat)&=U(\nhat+\nabla \phi)) \approx U(\nhat) + \nabla \phi \cdot \nabla U (\nhat) + \ldots
\end{align}
\label{eq:acc1}
\end{subequations}
Here we have truncated the expansion to first order in $\nabla \phi$, and which is referred to as the ``gradient approximation.'' 

\begin{figure}
  \centering  \includegraphics[width=0.5\textwidth]{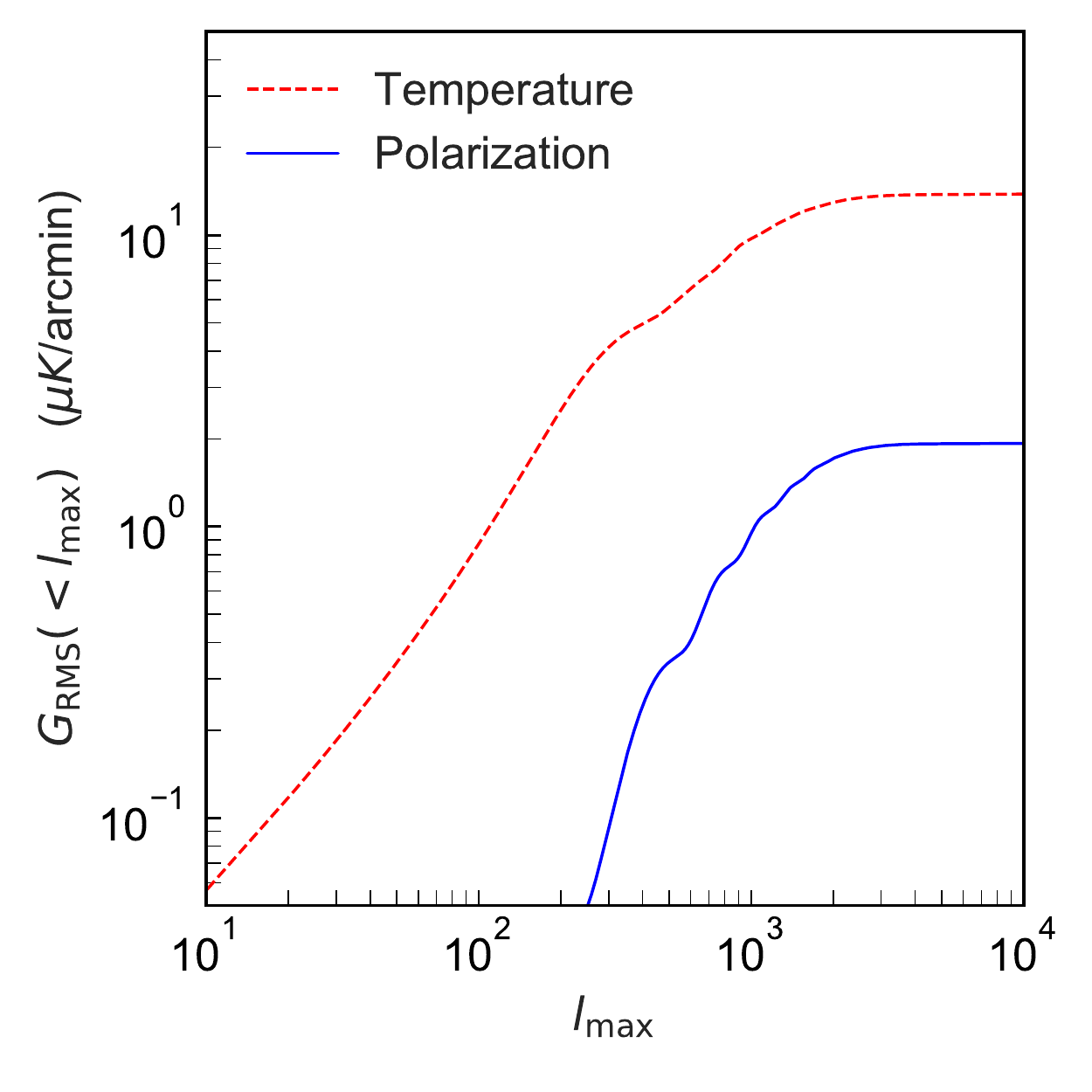}
    \caption{Gradient of the lensed temperature $T$ and polarization $E$ fields as a function of the filtering scale, $\ell_{\rm max}$, as defined in Eq \ref{eq:G_rms}. We note that the RMS gradient is almost a factor of 10 larger in temperature than in polarization, and this will make temperature lensing more sensitive to a fixed-mass lens than polarization on small scales (in absence of foregrounds).} 
    \label{fig_grad}
\end{figure}

Note that while the gradient approximation can be a poor approximation on large scales, it is  very good on small enough scales, for the following reason: CMB fluctuations have most of the power on large scales, small scales being suppressed by Silk damping. To quantify which scales contribute to the gradient, we can calculate the variance of the gradient when including multipoles $\ell$ only up to $\ell_{\rm max}$. For temperature, this is
\begin{equation}
G_{\rm rms}^2(<\ell_{\rm max})= \int^{\ell_{\rm max}}_{0} \frac{d^2 \bl}{(2\pi)^2} \ \ell^2C_\ell^{TT}
\label{eq:G_rms}.
\end{equation}
and a similar definition holds for polarization.  Figure \ref{fig_grad} shows $G_{\rm rms}$ as a function of $\ell_{\rm max}$, and we can see that the gradient variance entirely originates at $\ell \lesssim 2000$, as pointed out in \cite{hu2007cluster}. If we consider small lenses such that $\nabla \phi$ receives most of its contribution from $\ell > 2000$, the gradient approximation should be excellent, and in this paper we will study this regime, which is where analytic progress can be made.  For larger lenses, where this assumption fails, more expensive numerical maximum-likelihood methods should be employed to ensure optimality (\cite{hirata2003analyzing,carron2017maximum, Millea:2017fyd}).

As we have discussed in the introduction, the advantages of CMB lensing over galaxy lensing for measuring cluster masses become large at $z \gtrsim 1$, and we will show that most of those clusters are small enough in the sky for the gradient approximation to hold.  This makes cluster lensing an ideal application of our formalism.

For concreteness, in the rest of the paper we will work in terms of the temperature fluctuations $\Theta$, but the same formalism also applies to $Q$ and $U$, since they are deflected by the same vector.  Sometimes it will also be useful to work in terms of the lensing convergence, $\kappa$, defined as 
$\kappa(\nhat)=-\frac12 \nabla \cdot \d(\nhat) = - \frac12 \nabla^2 \phi(\nhat)$. We also note, that since the rms gradient of temperature fluctuations is almost an order of magnitude larger than that for polarization, temperature reconstruction is expected to be the dominant source of information on about small scales. Polarization, while statistically less powerful, is less affected by some of the foreground contamination and can provide useful consistency checks.

\subsection{Quadratic Estimator in the small-scale limit and Optimal Estimators}
\label{sec_QEvsOE}
CMB lensing introduces coupling between long and short wavelength modes and it is possible to construct a minimum-variance quadratic estimator for the lensing potential $\phi$. Note that the ``minimum-variance'' qualification here only applies to the class of estimators that are quadratic in the observed fluctuations, and is not a general statement of optimality. It has been previously shown that maximum likelihood or iterative methods can outperform the QE.  In this section we show that the QE is very suboptimal in the small-scale and low noise regime since it is subject to the cosmic variance on the gradient mode, while (in this regime) the particular realization of the gradient on the background of the lens can be measured without cosmic variance. In this section we will take the limit in which the reconstructed mode $L \gg \ell_{\rm Silk}$ (in practice $L \gtrsim 4000$), low noise and no-foreground limit.

The standard QE of Hu and Okamoto (\cite{hu2002mass}) for temperature can be written as\footnote{For compactness, we use the notation $\int_{\bl} \equiv \int \frac{d^2 \bl}{(2\pi)^2}$. Upper-case $\mathbf{L}$ denotes lensing multipole, while lower-case $\bl$ denotes temperature map multipole.}:
\begin{equation}
\hat{\phi}_{\rm QE}(\L) =  N(\L) \int_{\bl} \tTheta(\bl) \tTheta(\L - \bl) f(\bl, \L) 
\end{equation}
where the mode-coupling kernel is
\begin{equation}
f(\bl, \L) = \frac{ (\L - \bl) \cdot \L C^{TT}_{|\L - \bl|} + \bl \cdot \L C^{TT}_{\ell} }{ 2C^{\rm tot}_{\ell}  C^{\rm tot}_{|\L - \bl|} }\ \ .
\end{equation}
Here, $C_{\ell}^{\rm tot}$ includes contributions from instrumental noise, foregrounds, and the (lensed) primary CMB.  The reconstruction noise serves as the normalization in the estimator and represents the uncertainty in the reconstruction of $\phi(\L)$ due to chance correlations between different modes in an unlensed, Gaussian realization,
\begin{equation}
 N(\L)^{-1} = \int_\bl \frac{ \left[(\L - \bl) \cdot \L C^{TT}_{|\L - \bl|} + \bl \cdot \L C^{TT}_{\ell} \right]^2 }{ 2C^{\rm tot}_{\ell}  C^{\rm tot}_{|\L - \bl|} } \ \ \ \ .
\end{equation}

Taking the expectation value of this estimator over many CMB realizations for fixed $\phi$, this estimator recovers an unbiased mean deflection field (at first order in the lensing expansion). However, this reconstruction has significant noise for any given CMB realization.
This can be seen by taking the limit of reconstruction of small scale modes, with $L$ higher than where the primary CMB $C^{TT}_{\ell}$ has support, i.e. on scales much smaller than the Silk damping scale, $L \gg \ell_{\rm Silk}$. Since we will use $L$ for the reconstructed multipole and $\ell$ for the multipoles used in the reconstruction, we'll denote this limit by $L \gg \ell$.  We then have:
\begin{equation}
f(\bl, \L) \ \xrightarrow{L \gg \ell} \ 2 \ \frac{\bl \cdot \L C^{TT}_{\ell}}{2C^{\rm tot}_{\ell}  C^{\rm tot}_{L}},
\end{equation}
and
\begin{equation}
 N(\L)^{-1} \ \xrightarrow{L \gg \ell} \ 2 \int_\bl \frac{\left[ \bl \cdot \L C^{TT}_{\ell} \right]^2}{2C^{\rm tot}_{\ell}  C^{\rm tot}_{L}}.
\end{equation}
Therefore the quadratic estimator simplifies to\footnote{Technically, the convergence in the following equation would only be ``in probability'' and not at the field level, if $\Theta(\bl)$ and $\phi(\bl)$ were random variables with values that are uncorrelated for different $\bl$. In practice, the finite (and small) window function used to define the local gradient and nonlinear evolution of the potential make both continuous functions and the limit well defined. This subtlety doesn't affect any of our results.}
\begin{equation}
\hat{\phi}_{\rm QE}(\L) \longrightarrow \frac{\displaystyle \int_{\bl} \bl \cdot \L \ W_F(\ell) \tTheta(\bl) \tTheta(\L) }{\displaystyle \int_{\bl} (\bl \cdot \L)^2 \ W_F(\ell) C^{TT}_{\ell}} \ , \ \ \ \ \ {\rm with }\ W_F(\ell) = \frac{C^{TT}_{\ell}}{C^{\rm tot}_{\ell}} \ .
\end{equation}
In the low-noise, no foregrounds limit, we can take the Wiener filter $W_F(\ell) \rightarrow 1$ and note that 
\begin{equation}
\int_{\bl} (\bl \cdot \L)^2 \ W_F(\ell) C^{TT}_{\ell} \approx \int_{\bl} (\bl \cdot \L)^2 \ C^{TT}_{\ell} = L^2 \ \frac{1}{2} (\nabla \Theta)_{\rm rms}^2 \ .
\end{equation}
The quadratic estimator takes a simpler form of
\begin{equation}
\hat{\phi}_{\rm QE}(\L)\longrightarrow \frac{\tTheta(\L)}{L} \ \frac{\displaystyle  \int_{\bl} \bl \cdot \nhat_\L \ \tTheta(\bl) }{\frac{1}{2}(\nabla \Theta)_{\rm rms}^2}  \ ,
\end{equation}
where we have written $\L = L \ \nhat_\L$. Here $\int_{\bl} \bl \cdot \nhat_\L \ \tTheta(\bl) \approx (\nabla \Theta)_{\rm true,\nhat_\L}$ is the (realization dependent) measured background gradient projected in the direction parallel to $\L$.
We can then rewrite
\begin{equation}
\hat{\phi}_{\rm QE}(\L)\longrightarrow \frac{\tTheta(\L)}{L} \ \frac{(\nabla \Theta)_{\rm true,\nhat_\L} }{\frac{1}{2}(\nabla \Theta)_{\rm rms}^2}  \ .
\label{eq:sm-QE}
\end{equation}
This is the small-scale limit of the QE, which we will interpret soon. 

On the other hand, at $L > 4000$, the primary CMB is highly suppressed by diffusion damping and in our limit, all small scale fluctuation are created by lensing. In this regime, we can write the small-scale  fluctuations as
\begin{equation}
\tTheta(\nhat) \approx \nabla \phi \cdot \nabla \Theta(\nhat) \ .
\label{eq:grad_approx}
\end{equation}

Note that we can treat the locally measured gradient as constant on scales smaller than $1/\ell_{G}$, where $\ell_{G} \approx 2000$ is the multipole where the gradient becomes saturated as shown in figure \ref{fig_grad}. Then, there is a one-to-one correspondence between small scale lenses $\nabla \phi$ and measured fluctuations $\tTheta$. In Fourier space,
\begin{equation}
\begin{aligned}
\tTheta(\L) &= \int_{\bl} \bl \cdot (\L - \bl) \Theta(\bl) \phi(\L - \bl) \\ 
& \xrightarrow{L \gg \ell} \ L \phi(\L) \int_{\bl} \bl \cdot \nhat_\L \ \tTheta(\bl)  = L \phi(\L) \ (\nabla \Theta)_{\rm true,\nhat_\L}\ .
\end{aligned}
\end{equation}
We can call the solution of the previous equation for $\phi(\L)$ the ``Gradient Inversion'' (GI) solution:\footnote{The GI solution corresponds to the maximum likelihood solution in the limit considered here, where both the gradient and the short scale modes can be measured on this particular realization with $S/N \gg 1$. This can be seen by writing a likelihood based on Equation \ref{eq:grad_approx} and taking the noise to zero.}
\begin{equation}
\hat{\phi}_{\rm GI}(\L) \longrightarrow \frac{\tTheta(\L)}{L} \ \frac{1}{(\nabla \Theta)_{\rm true,\nhat_\L}} \ .
\label{eq:ML_est}
\end{equation}
Note that modes with $\L$ perpendicular to the gradient direction have $(\nabla \Theta)_{\rm true,\nhat_\L} = 0$ and cannot be reconstructed. This is because in this limit, lensing doesn't produce any effect perpendicular to the gradient direction and therefore no estimator can reconstruct such modes.

We can now compare the two estimators:
\begin{equation}
\begin{aligned}
L \ \hat{\phi}_{\rm QE}(\L) &\longrightarrow \tTheta(\L) \ \frac{ (\nabla \Theta)_{\rm true,\nhat_\L}}{\frac{1}{2}(\nabla \Theta)_{\rm rms}^2}, \\
L \ \hat{\phi}_{\rm GI}(\L) &\longrightarrow \tTheta(\L) \ \frac{ 1}{(\nabla \Theta)_{\rm true,\nhat_\L}} \ .
\end{aligned}
\end{equation}
Given a good enough experiment, $\tTheta(\L)$ and $(\nabla \Theta)_{\rm true,\nhat_\L}$ can be made measured to arbitrary accuracy, and therefore the error on $\hat{\phi}_{\rm GI}(\L)$ can be made arbitrarily small. Note that in this limit neither $\tTheta(\L)$ nor $(\nabla \Theta)_{\rm true,\nhat_\L}$ are random variables, and should therefore not be marginalized over.

This is not the case for the quadratic estimator, which can be rewritten as (in the limit of small noise), and identifying $\hat{\phi}_{\rm GI}$ with the ``true'' lensing potential $\phi_{\rm true}$,
\begin{equation}
\hat{\phi}_{\rm QE}(\L) = \hat{\phi}_{\rm GI}(\L) \ \frac{ (\nabla \Theta)_{\rm true,\nhat_\L}^2}{\frac{1}{2}(\nabla \Theta)_{\rm rms}^2} \longrightarrow \phi_{\rm true}(\L) \ \frac{ (\nabla \Theta)_{\rm true,\nhat_\L}^2}{\frac{1}{2}(\nabla \Theta)_{\rm rms}^2}
\end{equation}
Firstly, we notice that $\langle (\nabla \Theta)_{\rm true,\nhat_\L}^2 \rangle = \frac{1}{2} (\nabla \Theta)_{\rm rms}^2$, so that $\langle \hat{\phi}_{\rm QE} \rangle =\phi_{\rm true}$, where the expectation value is taken over realizations of the background gradient, for a fixed $\phi$. We see that while unbiased, the ``error'' that the QE makes is proportional to the difference between the (square) true gradient and the rms gradient. Since this quantity is on average of order the rms gradient itself, the fractional error of the QE is always of order unity per mode.

We can formalize the above intuition, showing that even with arbitrarily small experimental noise, there is a lower limit to the statistical error on the quadratic estimator $\sigma(\hat{\phi}_{\rm QE})$.
Defining 
\be
R^2_{\nabla}(\nhat_\L) = \frac{ (\nabla \Theta)_{\rm true,\nhat_\L}^2}{\frac{1}{2}(\nabla \Theta)_{\rm rms}^2} \ \ \ ,
\ee
we find that
\begin{equation}
\left ( \frac{\sigma^2(\hat{\phi}_{\rm QE})}{\phi_{\rm true}^2}\right )_{\rm min}(\L)  = \left\langle (R^2_{\nabla})^2 \right\rangle - \left\langle R^2_{\nabla} \right\rangle^2 = 3 -1 = 2
\end{equation}
In summary, the GI estimator can have arbitrarily low noise per mode for a good enough experiment, while the QE is limited to $S/N = 1/\sqrt{2}$ per mode (which is the same has having cosmic variance on the gradient mode).

\begin{figure*}
\centering  \includegraphics[trim={7cm 0 0cm 0},clip,width=1.0\textwidth]{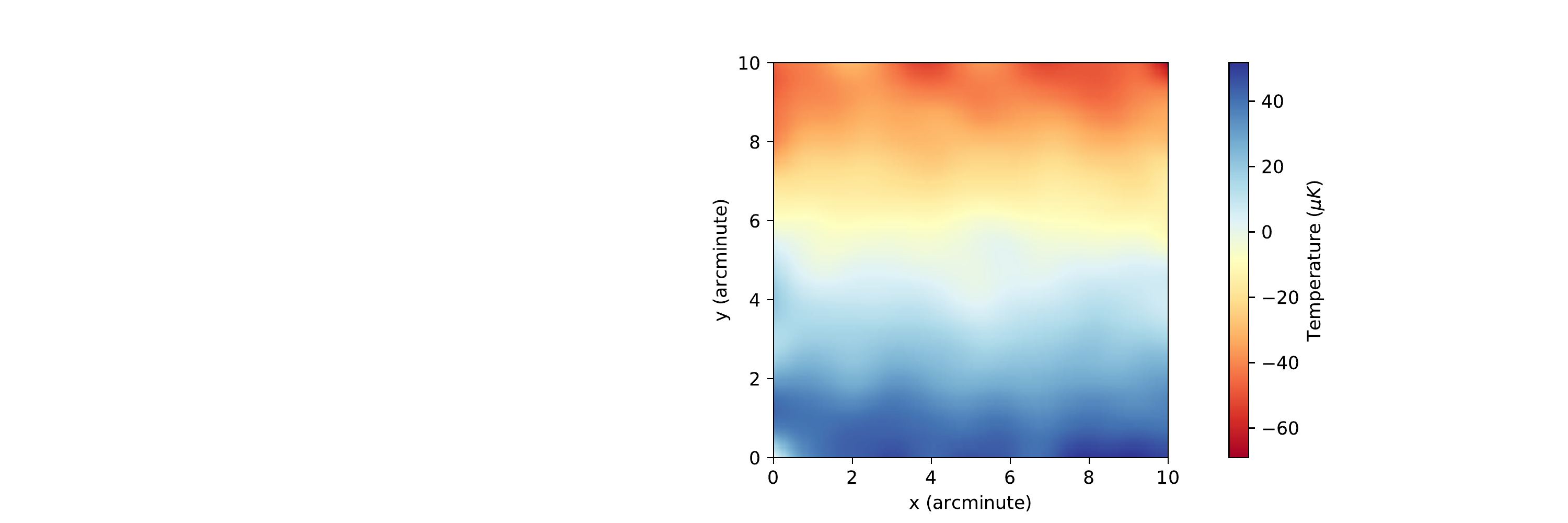}
  
  \centering  \includegraphics[trim={0.1cm 0 5cm 0},clip,width=1.0\textwidth]{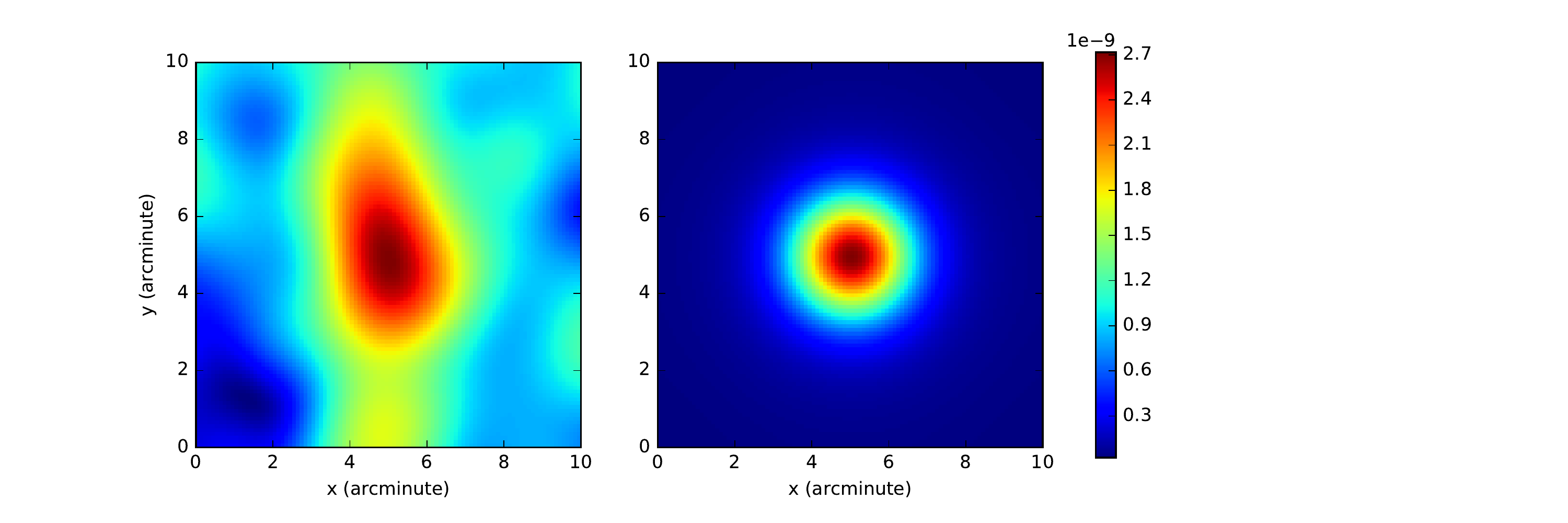}
  \centering 
  \includegraphics[width=0.40\textwidth]{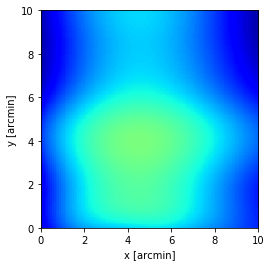}
    \caption{Top: Simulated lensed CMB Map of a constant gradient in the $\hat{y}$ direction, assuming $1$ arcmin beam and noise level of $1$ $\mu$K-arcmin. A cluster mass of $2 \times 10^{14} M_\odot$ at $z=2.0$ with an NFW profile (see Section \ref{nfw}) is present at the center. Middle Left: Reconstruction of the convergence map given the observed CMB map using the GI method.  Note that pure vertical modes (i.e. mass density with only $\hat{y}$ component) have been explicitly set to zero, while modes close to pure $\hat{y}$ are poorly reconstructed resulting in a residual vertical band. Middle Right: True convergence of cluster. Bottom: Quadratic estimator convergence of cluster at same color-scale.} 
    \label{reconst}
\end{figure*}

\subsection{One gradient, many independent modes?}
In this section we explore the covariance of the small scale lensing modes estimated using the quadratic estimator, in the very high-$L$, very low noise limit.  In this regime, we have shown that 
\be
\hat{\phi}_{\rm QE}(\L) \longrightarrow \phi_{\rm true}(\L) \ \frac{ (\nabla \Theta)_{\rm true,\nhat_\L}^2}{\frac{1}{2}(\nabla \Theta)_{\rm rms}^2} = \phi_{\rm true}(\L) \ R^2_{\nabla}(\nhat_\L)
\ee
Then the noise covariance between different small scale modes sharing the same background gradient becomes:
\be
N_{\rm QE}^{\phi \phi}(\L, \L') \equiv \langle (\hat{\phi}_{\rm QE}(\L) - \phi_{\rm true}(\L)) \ (\hat{\phi}_{\rm QE}(\L') - \phi_{\rm true}(\L')) \rangle_{\Theta}
\ee
where crucially the average is taken over realization of the long-wavelength CMB fluctuations, for a fixed large scale structure $\phi_{\rm true}$. 
we have
\be
\begin{aligned}
N_{\rm QE}^{\phi \phi}(\L, \L') & \longrightarrow \phi_{\rm true}(\L) \phi_{\rm true}(\L') \langle (R^2_\nabla(\nhat_\L) - 1)(R^2_\nabla(\nhat_{\L'}) - 1) \rangle \\
&= 2\cos^2(\zeta) \ \phi_{\rm true}(\L) \phi_{\rm true}(\L')
\end{aligned}
\ee
where $\zeta$ is the angle between $\L$ and $\L'$ and we have used $\left \langle R^2_\nabla(\nhat_\L) R^2_\nabla(\nhat_{\L'}) \right \rangle = 1 + 2\cos^2(\zeta)$ and  $\langle R^2_\nabla \rangle = 1$.

We conclude that for small scale modes not only the noise covariance matrix $N_{\rm QE}^{\phi \phi}(\L, \L')$ is not diagonal, but the off-diagonal terms become as large as the on-diagonal ones.  This means that when detector noise is negligible, modes sharing the same gradient become highly correlated, and treating them as independent (for example when forecasting cluster lensing constraints for future surveys), may lead to underestimation of the overall statistical noise.

We also note that this is a feature of the quadratic estimator, and won't be a limiting factor in a maximum-likelihood approach that includes information about the large-scale gradient. In principle, a maximum likelihood approach has no limit on the statistical significance that it can achieve\footnote{Further neglecting post-Born corrections and field rotation, as well as foregrounds and residual primary CMB.}.

\subsection{Example: Lensing Map Reconstruction}

In the previous section, we have derived a $\phi$ estimator for the pure gradient, low noise limit, which is found in Eq \ref{eq:ML_est}. Equivalently, we can rephrase that in terms of the lensing convergence $\kappa(\L) = - \frac{L^2}{2} \phi(\L)$, 

\begin{equation}
\hat{\kappa}_{\rm GI}(\L) \longrightarrow -\frac{L \tTheta(\L)}{2} \ \frac{1}{(\nabla \Theta)_{\rm true,\nhat_\L}} \ .
\end{equation}

In Figure \ref{reconst}, we show a comparison between the true and reconstructed convergence field, in the case of a massive lens with mass $2 \times 10^{14} M_{\odot}$ and following an NFW profile, as described in the next section.  We note that as expected, the modes with variation perpendicular to the gradient direction are not correctly reconstructed, explaining the vertical band in the reconstructed map. Those modes can be filtered out or downweighted in a real analysis since they have infinite variance.

In presence of finite noise, the error on the estimate of a given $\kappa(\L)$ noise, depends on the errors on both the large-scale gradient $(\nabla \Theta)_{\rm true,\nhat_\L}$ and the small scale temperature fluctuation $\tTheta(\L)$.

We expect the error on the gradient to be almost negligible in most cases, unless the gradient on the patch of interest happens to be much smaller than the rms, since most current or future CMB experiments should be close to cosmic variance limited at $\ell \lesssim 2000$,
\begin{equation}
\left(\frac{\sigma(\hat{\kappa}_{\rm GI})}{\kappa}\right)^2= \left(\frac{\sigma(\tTheta)}{\tTheta}\right)^2+\left(\frac{\sigma((\nabla \Theta)_{\rm true,\nhat_\L})}{(\nabla \Theta)_{\rm true,\nhat_\L}}\right)^2.
\end{equation}
Assuming that the errors are on the gradient and the small scales are independent. Note that the noise properties of $\hat{\kappa}_{\rm GI}$ are in general not trivial, since the noise is anisotropic (it depends on the angle between the gradient and $\L$), and the direction of the anisotropy depends on position on the sky. We defer a full treatment to future work, but we will discuss noise estimates in the specific case of cluster lensing in what follows.

\section{Measuring Cluster Masses with matched filtering}
\label{sec_mf}
A straightforward application of the formalism outlined above arises when the lens profile is known up to a normalization factor, such as when determining the mass $M$ of a cluster, lying on a constant gradient $\bi{G} = (\nabla \Theta)_{\rm true}$.  We have previously shown that the on small scales, the GI solution is linear in the measured temperature fluctuations, and therefore the minimum variance estimator for the overall amplitude of the lensing $A = M|G|$, can be obtained by a matched filter of the CMB map (\cite{haehnelt1996using}). Once both the amplitude $A$ and the background gradient $\bi{G}$ have been measured, the mass can be obtained by direct inversion $M \approx A / |G|$. 
In the context of cluster lensing, this was first investigated by (\cite{seljak2000lensing}), and here we revisit this point. 

\begin{figure}
  \centering  \includegraphics[width=0.45\textwidth]{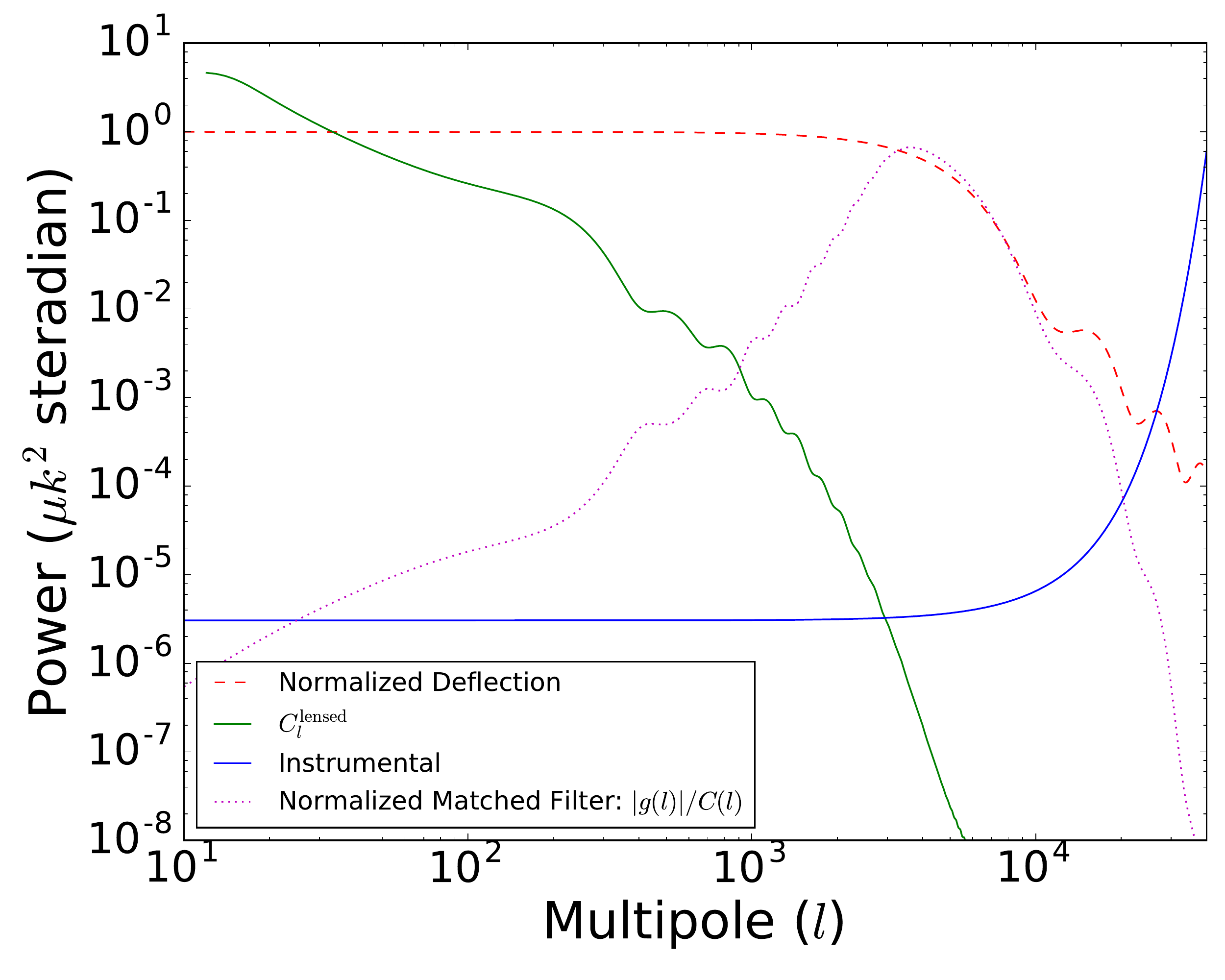}
    \caption{Components of matched filter in $\l$-space. At low $\l$ the cluster power will be suppressed by the primary CMB fluctuations and at high $\l$ it will be suppressed by the instrumental beam. Instrumental noise is for a 6 $\mu$K-arcmin experiment with a 1 arcmin beam. } 
    \label{fig_multipoles}
\end{figure}

A mass of a given profile on a pure gradient background CMB will create a dipole-like structure $g(\bi{\theta})$, aligned with the gradient direction, and with a free amplitude $A$ that we wish to measure. We then model the  total small-scale CMB anisotropy as
\begin{equation}
\tTheta (\bi{\theta})=Ag(\bi{\theta})+n(\bi{\theta})
\end{equation}
where $g(\bi{\theta})$ is the angular profile of the deflection angle of a known profile (i.e. NFW, Sersic, etc.) caused by the cluster, $A$ is an amplitude depending on the mass of the cluster and the CMB gradient, and $n(\bi{\theta})$ is noise, either from the instrument, foregrounds, or residual primary CMB. It is useful to work with the noise in Fourier space and its power spectrum,
\begin{equation}
\langle n(\bl) \ n(\bl')\rangle=(2 \pi)^2 C_\ell^{\rm tot}\delta(\bl+\bl').
\label{eq_noise}
\end{equation} 
The value of $A$ can be found by applying a linear filter $\Psi$ on the temperature map at the cluster position
\begin{equation}
\hat{A}=\int d^2\bi{\theta} \ \Psi(\bi{\theta})\tTheta(\bi{\theta}),
\label{A_est}
\end{equation}
where the optimal filtering function is given by (\cite{haehnelt1996using}),
\begin{equation}
\Psi(\bl) =\left[\int \frac{d^2\bl}{(2\pi)^2} \frac{|g(\bl)|^2}{ C_\l^{\rm tot}}  \right]^{-1} \frac{g(\bl)}{C_\ell^{\rm tot}}, 
\label{matched}
\end{equation}
can be found by minimizing the error on $\hat{A}$. We show components of our matched filter for a NFW profile in Figure \ref{fig_multipoles}.

The variance on $A$ is given by
\begin{equation}
\sigma^2_A=\left[\int \frac{d^2\bl}{(2\pi)^2} \frac{|g(\bl)|^2}{ C_\l^{\rm tot}}  \right]^{-1}
\label{sigma}
\end{equation}

Assuming an axi-symmetric cluster profile we can write
\begin{equation}
g(\bl)=g(\ell)\cos \phi_\ell 
\label{eq_deflection}
\end{equation}
where $\phi_\ell$ is the angle between the vector $\bl$ and the gradient direction.

Note that $\sigma_A \rightarrow 0$ as the noise power spectrum tends to zero, which is what we expect.

\subsection{NFW Profile}

\label{nfw}
For our work, we will assume that halos follow a NFW profile \cite{navarro1997universal}, with density given by 

\begin{equation}
\rho(r) \propto \frac{1}{r/r_s(1+r/r_s)^2},
\label{eq_nfw}
\end{equation}

with scale radius $r_s$ and normalization constant dependent on halo mass. The halo mass is related to the scale radius via the concentration parameter, $c$, which we will fix to 3.2 to allow direct comparison against \cite{hu2007cluster}. In practice, variation of the concentration and scale radius can be corrected for by combining CMB lensing information with other proxies for density, such as tSZ and X-ray observations. The lensing deflection profile (with the speed of light equal to one) for an NFW halo is
\begin{equation}
|\nabla \phi(\theta)| \propto g(\theta) = 16\pi G M \frac{D_{LS} D_{L}}{D_S} \frac{\rho_sr_s \theta_s}{\theta}h\left(\frac{\theta}{\theta_s}\right)
\label{eq_nfw_con}
\end{equation}
where, 
\begin{equation}
\rho_s = \frac{200}{3}\rho_{\rm crit} \frac{c^3}{\ln{c(1+c)}-c/(1+c)}
\label{eq_rhos}
\end{equation}
and $D_L$ is the comoving distance to the lensing, $D_S$ is the comoving distance to the source (i.e. the CMB), $D_{LS}$ is the comoving distance between the lens and the source, $\theta_s$ is the angle subtended by the scale radius and $M$ is the cluster mass. The functional dependence of the profile is 
\begin{equation}
    h(x)= 
\begin{cases}
     {\rm ln}(x/2) + \frac{2}{\sqrt{x^2-1}}{\rm arctan}\sqrt{\frac{x-1}{x+1}} ,  \quad & (x>1) \\
    {\rm ln}(x/2) + \frac{2}{\sqrt{1-x^2}}{\rm arctan}\sqrt{\frac{1-x}{1+x}} , \quad & (x<1) \\
    {\rm ln}(x/2) + 1\,,  \quad &(x=1)\,.
    \end{cases}
\label{eq_nfw_}
\end{equation}

\section{Idealized Example on CMB Map}
\label{sec_idealized}

To demonstrate this technique, we generated CMB realizations using HEALPix (\cite{healpix}\footnote{http://healpix.sourceforge.net}), with $N_{\rm side}$ = 2048, from which we extracted $20'$ square cutouts. For these images we use the lensed CMB power spectra to include the effects of lensing by large scale structure.  The effect of lensing from an NFW profile with a given concentration and mass is then added, together with detector noise, as shown in Figure \ref{fig_sims2x2}. We then reconstruct the mass using the matched filter technique described above, the first step being a measurement of the average the background gradient, as discussed below.

\subsection{Reconstruction of the background gradient}
\label{subsec_recon}
Correct determination of the cluster mass depends on being able to accurately extract the mean background gradient \cite{holder2004gravitational}, without bias from the cluster lensing signal or deviation from the pure gradient approximation. Fortunately, as discussed in section \ref{sec_effects}, this gradient has little variation on cluster scales. We can define the average gradient on a small patch centered on the cluster $i$ as $\bi{G}_i = [(\nabla \Theta)_{\rm true}]_{{\rm patch\ } i}$, so that
\begin{equation}
\bi{G}_{i}= \frac{1}{C}\int_{\ell < 2000} \frac{d^2 \bl}{(2 \pi)^2} \ [\bl \ \tTheta(\bl)]_{{\rm patch\ } i}
\label{G_av}
\end{equation}
where $C$ is proportional to the area being integrated over. One should be careful to not average over too large of a patch such that variations in the primary CMB would become a concern. For our analysis we find that a $8'$ box around the center of the cluster works adequately.

This measure is robust even in the presence of noise, resulting in less than a 0.1\% error in gradient extraction for an experiment with a 1 arcmin beam and 1 $\mu K$-arcmin sensitivity, compared to the ideal case. The error scales roughly linearly with instrumental sensitivity and has very small dependence on lensing halo mass and redshift. 

\begin{figure}
  \centering  \includegraphics[width=0.45\textwidth]{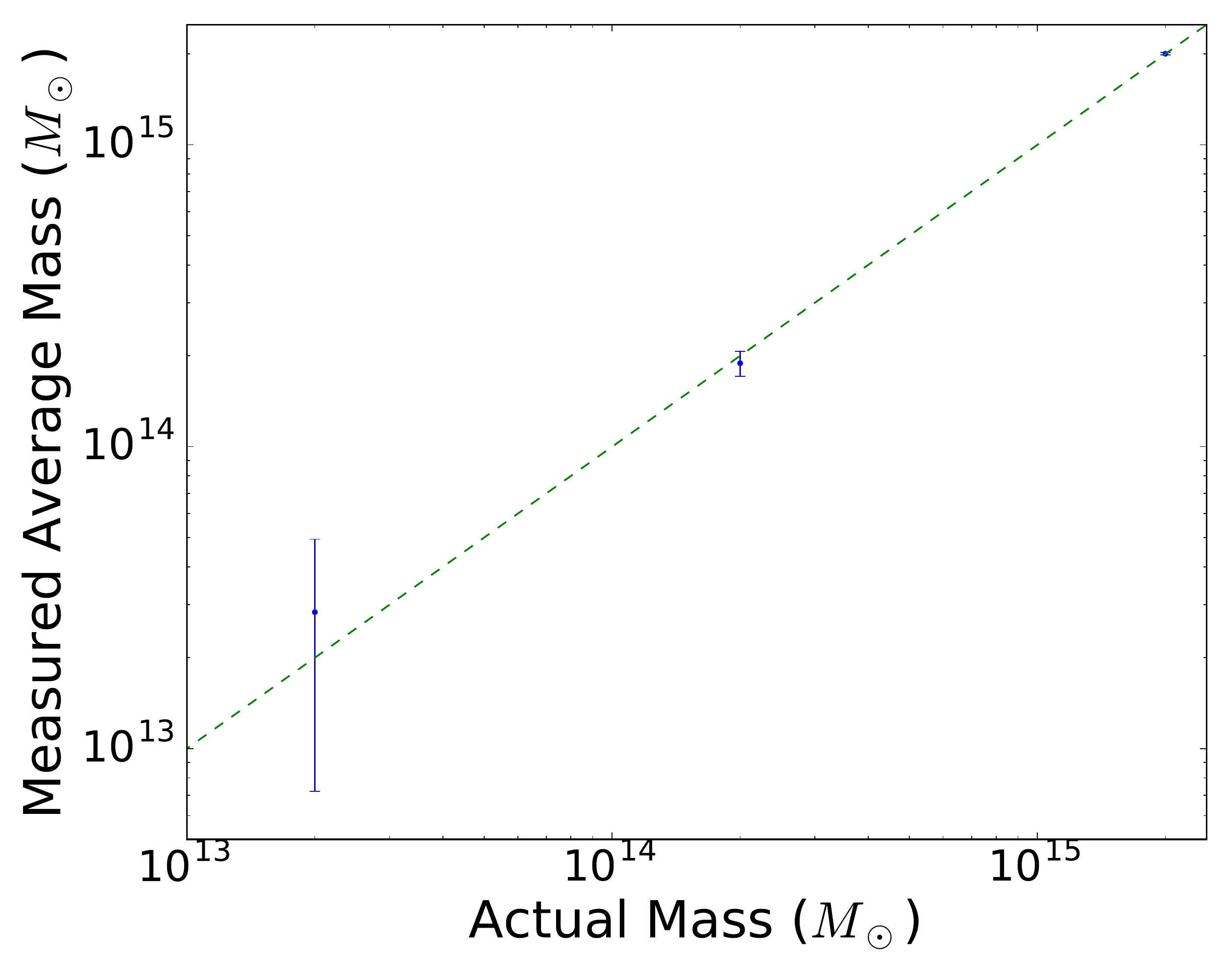}
    \caption{Cluster mass reconstruction as a function of mass for our 1000 realistic lensed CMB realizations. We have fixed $z=0.7$, $c=3.2$, $\Delta_T=6 \mu$K-arcmin, and beam FWHM = 1 arcmin. Our reconstructed mass has no measurable bias.} 
    \label{fig_sims}
\end{figure}

In addition, the presence of clusters does not strongly bias the measurement of the gradient itself. This shown explicitly in Figure \ref{fig_sims}, where we simulated the lensing of the CMB by massive clusters, extracted the gradient from those lensed images, and applied the matched filter prescription to measure the mass. As shown, there is no significant bias in the measured mass as it relates to the real mass.

\subsection{Measuring the lens mass}
\label{sec_measuremass}

Here we briefly summarize the procedure used for estimating the mass of simulated clusters. For a given simulated lensed image, we perform the following procedure:
\begin{enumerate}
\item Find the direction and magnitude of the gradient $\bi{G}_i$ at the center of the cluster, as in Equation \ref{G_av}.
\item Choose the axis of the matched filter (i.e. the $\phi_\ell$ = 0 direction), aligned with the direction of the gradient. This is important due to the $\cos \phi_\ell$ term in Equation \ref{eq_deflection} establishing the antisymmetry of the filter in real space.
\item Apply the matched filter defined in Equation \ref{matched} and perform the integral in Equation \ref{A_est} to obtain an amplitude $\hat{A}_i$.

\item Since $A_i = |G_i| M_i$ when correctly normalized, the mass $M_i$ can be estimated as $\hat{M}_i = \hat{A}_i /|G_i|$, with error dependent on the local size of the gradient as expected: $\sigma(M_i) = \sigma_A / |G_i|$. For a sample of many different clusters, the mean mass can be obtained by inverse noise weighting each measurement as explained in the next section.
\end{enumerate}

\subsection{Gradient-based weighting of samples}
\label{sec_gradweight}

\begin{figure*}

  \centering  \includegraphics[width=0.95\textwidth]{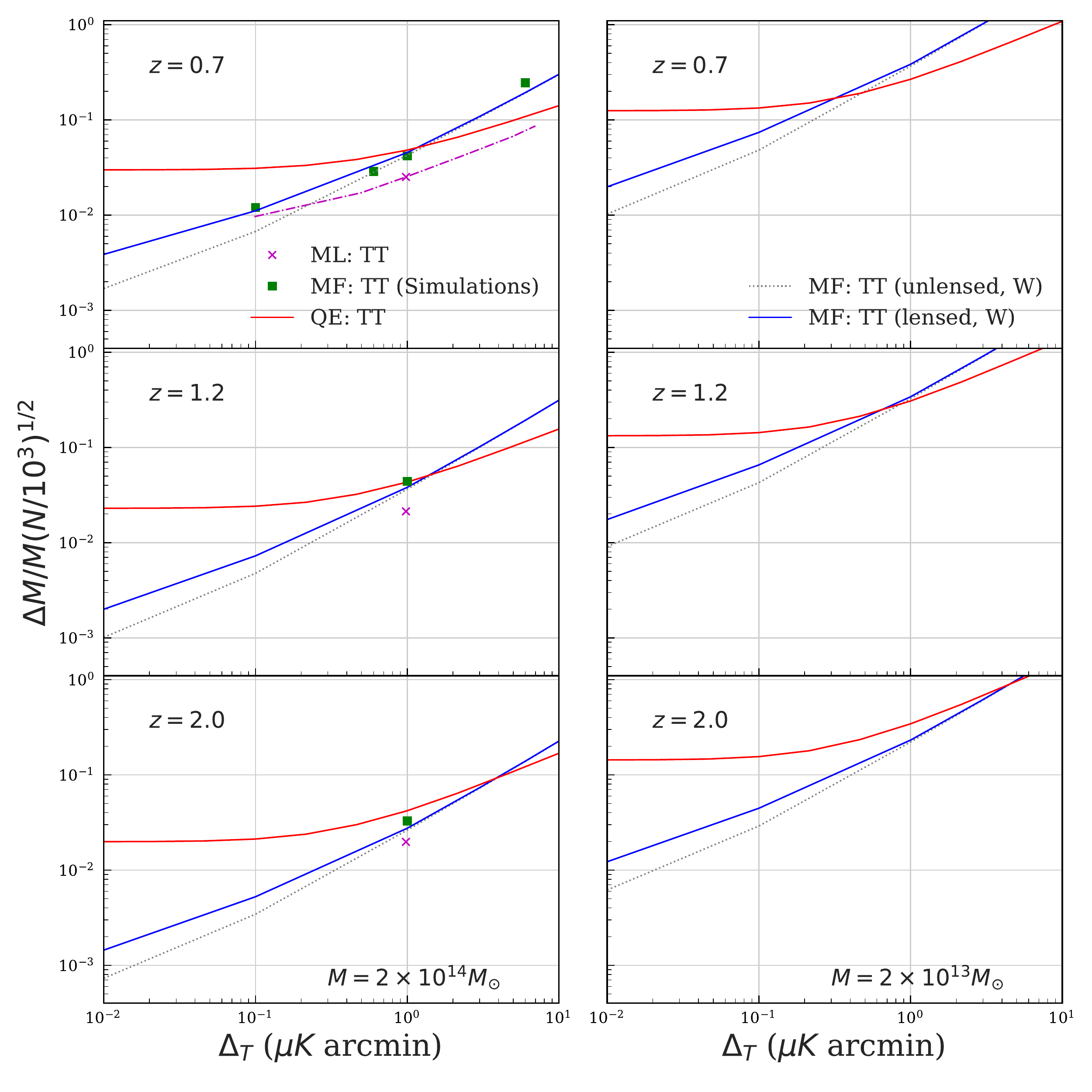}
    \caption{Mass sensitivity assuming 1 arcmin beam and no foregrounds using different techniques for redshifts $z=0.3,0.7,$ and $1.2$, as well as for two different masses, $2 \times 10^{14} M_{\odot}$ (left) and $2 \times 10^{13} M_{\odot}$ (right). In addition to the matched filter result from this work (MF) we show the quadratic estimator (QE) from \protect\cite{hu2007cluster}, which has an essentially flat redshift dependence at $z>0.5$ for a 1 arcmin beam (see also Figure 3 in \protect\cite{melin2015measuring}). We have also shown the maximum likelihood (ML) TT result from \protect\cite{raghunathan2017measuring} (see Figure 2). In the middle panel, we show the effect of using the lensed vs unlensed CMB as a source of noise. For all $\l$ integrals we use an $\l_{\rm min}=10$ and $\l_{\rm max}=40,000$ (beyond the beam cutoff scale). The QE outperforms the MF for large values of the noise, since in this regime most of the contribution to the $S/N$ will come from modes with $\ell < 4000$, where the MF is no longer optimal.
} 
    \label{fig_3x1}

\end{figure*}

Assume that we have a collection of $N$ clusters all of the same mass, $M$, but each in their own background gradient, $G_i$, which we assume is extracted with no noise from observations. Then the matched filter output for each cluster $i$, $\hat{A}_i$ (from Equation \ref{A_est}), and it's uncertainty, $\sigma_{A}$ (which is approximately independent of the cluster mass in a uniform sample, and only determined by detector noise and foregrounds), are given by 
\begin{align}
\hat{A}_i &= |G_i| M_i, \\
\sigma_{A} &= |G_i| \sigma(M_i).
\label{sigma_A}
\end{align}
The minimum variance estimate of the mean mass is obtained by  inverse noise weighing individual measurements. Noting that from Equation \ref{sigma_A}, $\sigma(M_i)$ is inversely proportional to $|G_i|$, we have
\begin{equation}
\hat{M} = \frac{\sum_i \frac{|G_i| \hat{A}_i}{\sigma_{A}^2}}{\sum_i \frac{|G_i|^2}{\sigma_{A}^2}} 
\label{sample_A_i}.
\end{equation}
Where the uncertainty on the mean mass is
\begin{equation}
\sigma^2(\hat{M}) = \left [ \sum_i \frac{|G_i|^2}{\sigma_{A}^2} \right ]^{-1} = \frac{1}{N} \frac{\sigma_{A}^2}{G_{\rm rms}^2}
\label{sigma_A_i} \ \ ,
\end{equation}
The latter equality being true in the limit of a large sample size. Equation \ref{sigma_A_i}, together with the matched filter error $\sigma_A$ given by Equation \ref{sigma}, provides the basis for our results in Figure \ref{fig_3x1}. Note that uniform weighting would lead to an increase of the mass uncertainty.

\subsection{Results}

In Figure \ref{fig_3x1} we show the analytical calculation of the mass uncertainty  using the formalism of Equation \ref{sigma}, allowing the integral to be cut off by the beam (see Figure \ref{fig_multipoles}). Our fiducial case include the residual \textit{lensed} CMB as a source of noise in $C_\ell^{\rm tot}$, while we also show the case in which the unlensed CMB is used instead. In practice, the fiducial case is conservative in the sense that no delensing is assumed, while any amount of delensing (either internal (\cite{carron2017internal}) or with the use of other tracers (\cite{smith2012delensing})) will place the noise between our two extreme cases. 


For comparison, we show the performance of the quadratic estimator (QE) of \cite{hu2007cluster}, as well as the maximum likelihood (ML) result of \cite{raghunathan2017measuring}. For the QE, we use the formalism of \cite{hu2007cluster} (see Equation 30 there) and we integrate over $\ell$ well past the beam cutoff scale ($\ell \approx 10,000$) for a fair comparison between the methods. The matched filter outperforms the quadratic estimator in the low noise limit as well as the high redshift limit. For a cluster of a constant mass, its apparent size decreases with increasing redshift and the background gradient is closer to a pure gradient where the match filter technique is optimal.

As expected by the analytical calculations in Section \ref{sec_QEvsOE}, the quadratic estimator performance saturates at signal to noise ratio $S/N \sim 1$ per cluster. Meanwhile, the matched filter keeps improving for small values of $\Delta_T$. 

As a check, we validate our error estimate on simulations. To do this, we have taken 1,000 cutouts from full-sky realizations of the lensed primary CMB, as described in Section \ref{sec_idealized} and placed an NFW cluster with constant mass and $c=3.2$ at the center of the image and lensed the background primary CMB. We then measured the gradient as described in Section \ref{subsec_recon}, and found their mass using the procedure outlined in Section \ref{sec_measuremass}. We then weight the clusters according to their background gradient, as described in \ref{sec_gradweight}, to find the average measured mass; the results of which are shown in Figure \ref{fig_3x1}. 

\subsection{Effect of Beam Size}

\begin{figure}
  \centering  \includegraphics[width=0.50\textwidth]{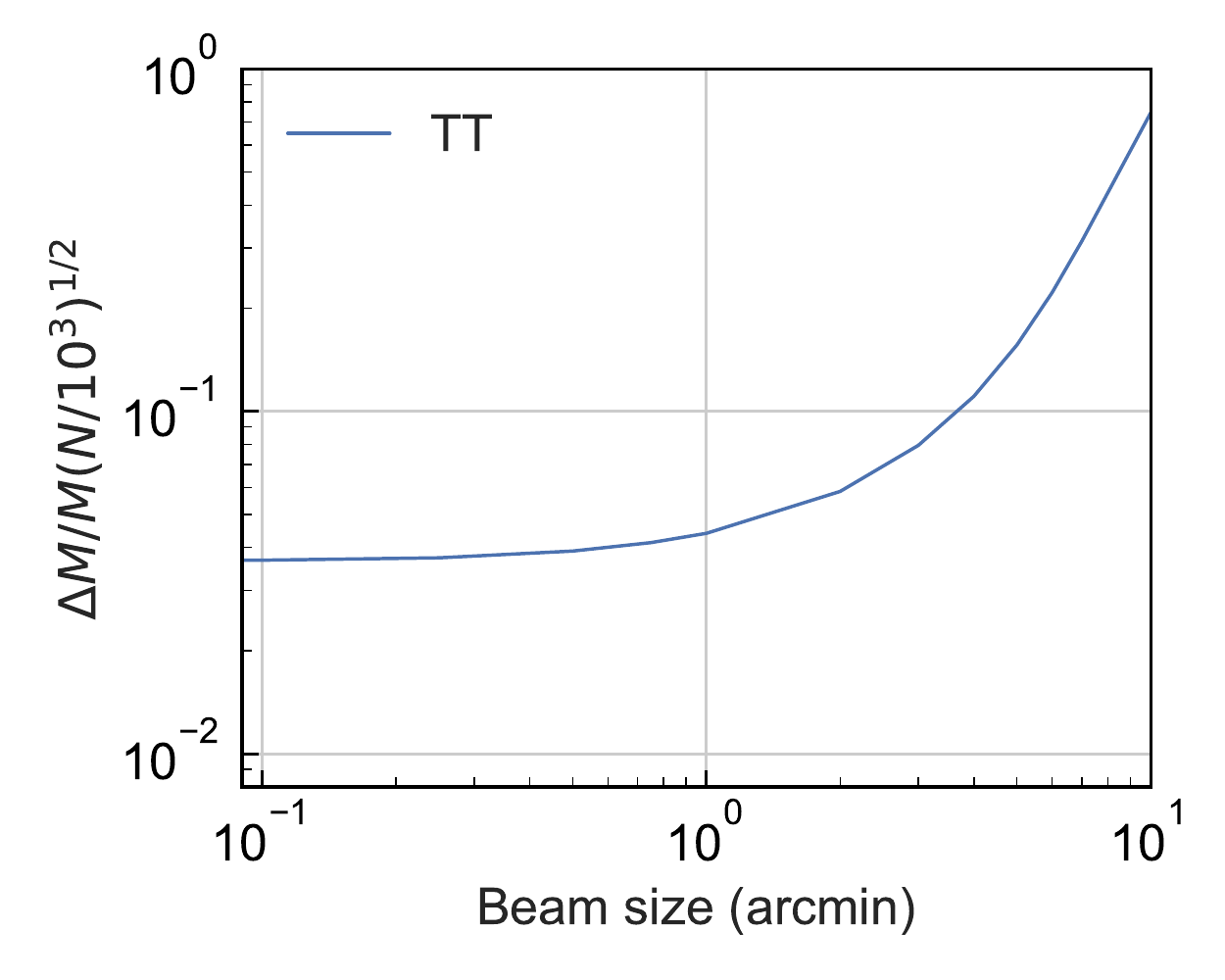}
    \caption{Effect of beam size on ability to extract mass, assuming 1 $\mu$K-arcmin instrumental noise, a $2 \times 10^{14} M_\odot$ cluster at $z=0.7$, and a $N=1000$ sample. At $0.3$ arcmin effectively all the power of an NFW cluster is captured in the matched filter and no more improvements are possible by decreasing beam size.} 
    \label{fig_beam}
\end{figure}

We show the effect of beam size in Figure \ref{fig_beam}. As beam size decreases, less of the power from the deflection angle is cut off (see Figure \ref{fig_multipoles}), and there is a significant improvement when going to higher resolution, up to about 1 arcmin beam (FWHM) for a $2 \times 10^{14} M_\odot$ cluster at $z=0.7$. However, when the cluster is well resolved, the improvements become marginal. Smaller beam size may be useful in extracting other information from the cluster, such as concentration or other details about the profile.

\subsection{Effect of Foreground Emission and other Secondary Anisotropies}

Here we briefly comment on the impact of foregrounds on the matched filter estimator and we leave a full exploration with simulation and correlated emission to future work. We note that we expect many of the issues outlined in the detailed study of \cite{raghunathan2017measuring} in the context of a maximum likelihood approach to be relevant here as well, with some caveats that we now explain.

Firstly, the effects of residual foreground emissions uncorrelated with the cluster in question and with known power spectra can be incorporated in this formalism as an additional source of noise contribution in Equation \ref{eq_noise}. In this case, the noise appearing in the plots would refer to ``effective noise'' after component separation, and we note that most of foregrounds can be in part mitigated by a multi-frequency analysis.

An exception is the kinematic Sunyaev-Zel'dovich (kSZ) effect caused by the bulk motion of free electrons (\cite{Sunyaev:1970eu}). Unlike other foregrounds, the kSZ effect cannot be subtracted out by using multifrequency observations as it preserves the black body spectrum of the CMB, and therefore it represents an additional source of noise that is hard to overcome. As the measurements of the kSZ effect improve, it may be possible to construct a template of the emission and reduce the residual kSZ noise by a factor of order unity.

Any emission from the cluster itself, or from matter correlated with it should also only amount to extra noise, and not a bias, unless this emission is correlated with our estimate of the background gradient of the unlensed CMB. This is because the matched filter only detects dipole contributions aligned with the background gradient, due to the $\cos \phi_\ell$ term in Equation \ref{eq_deflection}.

While not technically a foreground, gravitational lensing from uncorrelated structures in the universe produce extra small-scale anisotropy which again acts as a source of noise, which is  included in the fiducial analysis throughout the paper.
In Figure \ref{fig_3x1}, we show that the difference in error caused by using the lensed vs unlensed CMB power spectra is only relevant at small noise levels. The effect of matter correlated with the halo is to enhance the lensing signal by what is usually known as the ``two halo term''.  In a real analysis, the whole matter profile would be obtained and the one and two halo contributions would be fit simultaneously, just as performed in most galaxy-galaxy lensing analyses.
 
\section{Relation to previous work}
\label{sec_comparison}
Here we briefly compare the match filtered technique with other work in literature. Direct matched filtering techniques were first discussed as applicable to CMB cluster lensing in (\cite{seljak2000lensing}), and extensions in (\cite{holder2004gravitational, vale2004cluster, Dodelson:2004as}). Most of the subsequent work has focused on the use of quadratic estimators, as modified by (\cite{hu2007cluster}) to ensure unbiased results (\cite{melin2015measuring, Maturi:2004zj}) This and other works based on the QE suffer from the statistical limitations discussed in our paper.=

Iterative techniques based on the quadratic estimator have been introduced in (\cite{yoo2008improved,yoo2010lensing}). In the limit of large number of iterations the result should converge to the maximum likelihood solution in a unbiased way. 

A maximum likelihood framework for cluster masses was developed in (\cite{lewis2006cluster, baxter2015measurement}) and further studied and extended to polarization in (\cite{raghunathan2017measuring}), which also provides extensive discussion of possible contamination from foregrounds. While it is not possible to outperform a full maximum likelihood approach, Figure \ref{fig_3x1}, shows that for high redshift clusters or small noise limit, the matched filter presented in this paper approaches the maximum likelihood performance, but with only a small fraction of the computational cost.

On the experimental front, cluster lensing has first been detected using the quadratic estimator by (\cite{madhavacheril2015evidence}) using data from the ACTPol experiment and CMASS galaxies from the BOSS survey. Subsequently, it has also been detected by the SPT collaboration using a Maximum-Likelihood technique (\cite{baxter2015measurement, Baxter:2017ixz}), as well as by the Planck collaboration (\cite{Ade:2015fva}).

Recently there has been work towards implementing a full maximum likelihood lensing map reconstruction without any assumptions about the shape of the lensing potential (\cite{carron2017maximum}), based on previous work (\cite{hirata2003reconstruction}). Going forward, further developing these numerical techniques and applying them to small scale lenses would be very useful, but it currently appears challenging to obtain rapid convergence on small scales.

\section{Summary}
\label{sec_sum}

In this work we have explored the limitations of quadratic estimator reconstruction of CMB lensing from Temperature, in the small scale, low noise limit. We have shown that the quadratic estimator is fundamentally limited by suboptimal weighting, while in no-foregrounds, small noise limit, small lenses should be measurable arbitrarily well. This problem can be overcome by using different techniques, such as direct gradient inversion presented in this paper, or using more computationally expensive maximum likelihood methods.

An obvious application of the direct gradient inversion is reconstruction of cluster masses from  CMB lensing. Once a profile for the potential is assumed, optimal weighting of the reconstructed modes is equivalent to a matched filter that we have explored in this work and which out-performs the quadratic estimator over a wide range of instrumental sensitivities and redshifts; focused primarily on the low noise and small angular scale limit. The matched filtered approach relies on the antisymmetric nature of lensing as well as the separation of scales between cluster-lensing and the variation in the primary CMB. The antisymmetric nature of the matched filter is a useful feature which allows it to be less sensitive to kSZ and other foregrounds coming from the cluster.

The matched filter studied here can be extended to fit additional parameters such as the cluster concentration (see for example (\cite{Dodelson:2004as})) or the amplitude of the two-halo term. Alternatively, tight priors can be put on them by studying the halo clustering, or detailed tSZ/X-ray observations. The effect of miscentering between the observed and true halo center needs to be quantified since it will lead to a bias if not properly accounted for, as pointed out by (\cite{raghunathan2017measuring}). At the same time, correlation between the selection function and other observables can produce biases in the inferred mean mass for the sample. For example, if halos elongated along the line of sight are preferentially selected, the inferred mass is likely to be overestimated due to the alignment of the halo with the surrounding cosmic structure. Detailed simulations and modeling are required to estimate the size of these effects for the particular sample being analyzed.

The technique discussed in this paper outperforms the QE for the noise level expected of future CMB experiments, and is computationally very efficient. 
Moreover, the noise calculation is analytically tractable, allowing direct forecasting and Fisher-type analysis without having to simulate a large number of clusters. 

\section*{Acknowledgments}
We appreciate helpful discussions with Eric Baxter, Anthony Challinor, Shirley Ho, Wayne Hu, Mathew Madhavacheril, Srinivasan Raghunathan, Emmanuel Schaan, Neelima Sehgal, Uro${\rm \check{s}}$ Seljak, Kendrick Smith, and David Spergel. 
BH is supported by the NSF Graduate Research Fellowship, award number DGE 1106400. SF thanks the Miller Institute for Basic Research in Science at the University of California, Berkeley for support. BDS acknowledges support from an STFC Ernest Rutherford Fellowship.

This research used resources of the National Energy Research Scientific Computing Center, a DOE Office of Science User Facility supported by the Office of Science of the U.S. Department of Energy under Contract No. DE-AC02-05CH11231.

\bibliographystyle{mnras}

\bibliography{main}

\end{document}